\documentclass[12pt,a4paper]{article}

\usepackage{setspace}
\usepackage[usenames,dvipsnames,svgnames,table,x11names]{xcolor}
\definecolor{OxfordBlue}{RGB}{0,62,133}
\definecolor{GreenM-A}{RGB}{255,0,0}
\definecolor{Green2M-A}{RGB}{255,0,0}
\definecolor{detail}{RGB}{255,0,0}
\usepackage[top=1in, bottom=1in, left=0.75in, right=0.75in]{geometry}
\usepackage{amsmath}
\usepackage{braket}
\usepackage{graphicx}
\usepackage{amsfonts}
\usepackage{mathrsfs}
\usepackage{float}
\usepackage{amssymb}
\usepackage{booktabs}
\usepackage{changepage}
\usepackage{tabu}
\usepackage{pbox}
\usepackage[colorinlistoftodos]{todonotes}
\usepackage{tikz} \usetikzlibrary{shapes}
\usepackage{enumitem}
\usepackage{upgreek}
\usepackage{datetime} \settimeformat{ampmtime}
\usepackage{simpler-wick}
\usepackage{ulem}
\usepackage[new]{old-arrows}
\usepackage{bm}
\usepackage[colorlinks=true, linkcolor=OxfordBlue, citecolor=OxfordBlue, filecolor=OxfordBlue, urlcolor=OxfordBlue, linktocpage=true]{hyperref}
\usepackage{cite}
\usepackage[utf8]{inputenc}

\newif\ifcomments
\commentsfalse 

\newif\ifdetails
\detailsfalse 

\newcounter{eg}[section]

\newcommand{\no}[1]{\textbf{:}#1\textbf{:}}

\newcommand{\dd}{{\rm d}}

\newcommand{\widesim}{\raisebox{-4pt}{$\widetilde{\qquad}$}}

\begin{document}

\title{
{Superconformal algebras for twisted connected sums \\ and $G_2$ mirror symmetry}
}

\author{Marc-Antoine Fiset}

\date{\vspace{-5ex}}
\maketitle
\thispagestyle{empty}
\begin{center}
\textit{ \small
Mathematical Institute, University of Oxford\\Andrew Wiles Building, Woodstock Road\\Oxford, OX2 6GG, United Kingdom
}\\\vspace{5pt}
\href{mailto:marc-antoine.fiset@maths.ox.ac.uk}{\texttt{marc-antoine.fiset@maths.ox.ac.uk}}\\\vspace{10pt}
\end{center}
%
%

\abstract{We realise the Shatashvili--Vafa superconformal algebra for $G_2$ string compactifications by combining Odake and free conformal algebras following closely the recent mathematical construction of twisted connected sum $G_2$ holonomy manifolds. By considering automorphisms of this realisation, we identify stringy analogues of two mirror maps proposed by Braun and Del Zotto for these manifolds.}


\newpage

\tableofcontents

\vspace{10pt}

\section{Introduction}

The recent construction \cite{MR2024648,MR3109862,Corti:2012kd} of millions of examples of so-called twisted connected sum (TCS) compact $7$-dimensional manifolds with holonomy group $G_2$ sparked much interest in the physics community in particular for their role in supersymmetric compactifications \cite{Halverson:2014tya, Halverson:2015vta, Braun:2016igl, Guio:2017zfn, Braun:2017ryx, Braun:2017csz, Braun:2017uku, Braun:2018joh, Braun:2018fdp}. The previously known constructions of $G_2$ manifolds have on the other hand inspired the development of $2$-dimensional superconformal algebras (SCA) particularly well tailored to investigate stringy properties of these geometries. Joyce's original construction \cite{MR1424428, Joyce2000}, based on torus orbifolds, directly inspired the work of Shatashvili and Vafa \cite{Shatashvili:1994zw} which led to a surge of interest in this subject \cite{Shatashvili:1994zw, Papadopoulos:1995da, Becker:1996ay, Acharya:1996fx, Figueroa-OFarrill:1996tnk, Acharya:1997rh, Blumenhagen:2001jb, Gukov:2002jv, Roiban:2002iv, Eguchi:2001ip, Noyvert:2002mc, Sugiyama:2001qh, Sugiyama:2002ag, Eguchi:2003yy, Gaberdiel:2004vx, deBoer:2005pt}. In particular they highlighted the higher spin algebra governing supersymmetric type II compactifications on $G_2$ holonomy manifolds. We will refer to it as the Shatashvili--Vafa $G_2$ SCA. Arguing from this algebra, conjectures were made in \cite{Shatashvili:1994zw} about analogues of mirror symmetry and topological twists for exceptional geometries.

Another possible approach to constructing $G_2$ manifolds is to start with the product of a Calabi--Yau $3$-fold with a circle and then take the quotient by an involution (and eventually desingularise) \cite{MR1424428}. $G_2$ manifolds admitting such a construction are also closely related to particular superconformal algebras, as first described in \cite{Figueroa-OFarrill:1996tnk}. More precisely, the $G_2$ SCA was shown to be realisable in terms of free fields and the higher spin SCA due to Odake \cite{Odake:1988bh} known to be related to the holonomy group $SU(3)$.

In this note we combine these elements and introduce the superconformal algebra appropriate for twisted connected sum $G_2$ holonomy manifolds. More precisely, we prove that the Shatashvili--Vafa $G_2$ algebra can be obtained from copies of Odake algebras ($\text{Od}^n$) and free fields in a way that mimics the TCS construction. A key step is to match the two halves of the construction with an algebra automorphism $F$ as explained more fully in the vicinity of equation~\eqref{eq:resultForIntro},
\begin{equation}
G_2\,\text{SCA} \longhookrightarrow \frac{\text{Od}^{n=2} \times (\text{Free})^3}{F} ~~ \Big/ ~ \langle N^1 , N^2\rangle \,.
\end{equation}
(Here $N^1$ and $N^2$ refer to an ideal modulo which our statements hold.) At its heart, our contribution is an explicit computation of all operator product expansions (OPE) underlying a series of vertex operator algebraic statements. We thus expect it to find interest in mathematical contexts.

The clear relationship of our result with TCS manifolds calls readily for mirror symmetry applications in type II string theory, in the sense recently suggested by Braun and Del Zotto \cite{Braun:2017ryx,Braun:2017csz}. With this in mind, we describe in section~\ref{sec:AutomTCS} two consistent automorphisms of our algebraic construction. We propose to regard these as conformal field theoretic versions of the $G_2$ mirror maps $\mathcal{T}_4$ and $\mathcal{T}_3$ that were argued for from a geometrical point of view in \cite{Braun:2017csz}. In this sense, we expect our results to help venture away from the geometrical locus of type II string compactifications on TCS manifolds.

The rest of this note is organised as follows. We cover aspects of the TCS construction in section~\ref{sec:TCSgeometry}. In section~\ref{sec:SCAandG2}, we briefly review the $G_2$ SCA. Generic comments about conformal algebras and null ideals are made in appendix~\ref{app:algebras}. We also provide in this appendix all OPEs needed in the main calculation. We exhibit our TCS algebra realisation in section~\ref{sec:Realisations}. Automorphisms of the realisation and mirror symmetry are discussed in section~\ref{sec:Automorphisms}. We close in section~\ref{sec:Conclusion} with a fuller discussion of our results.

\subsection{TCS geometry} \label{sec:TCSgeometry}

We begin with a short informal account of the twisted connected sum (TCS) construction of $7$-dimensional manifolds with holonomy group $G_2$. This technique was pioneered by Donaldson and Kovalev \cite{MR2024648} and substantially improved and developped by Corti, Haskins, Nordstr\"om and Pacini \cite{MR3109862,Corti:2012kd}. We refer to the work of these authors for details. For basics about $G_2$-structure geometry, we recommend \cite{Joyce2000,Grigorian:2009ge,Corti:2012kd}.

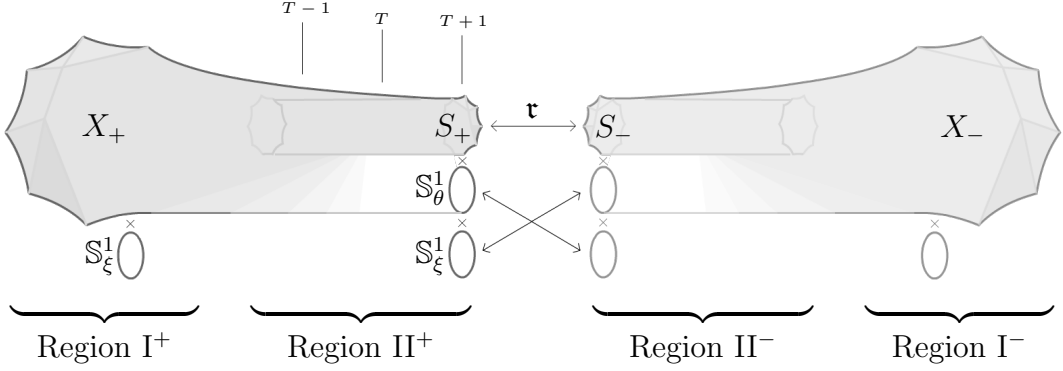
\begin{figure}[htbp]
\begin{center}
\begin{tikzpicture}[scale=0.9]
\node at (0,0) {\includegraphics[scale=0.252]{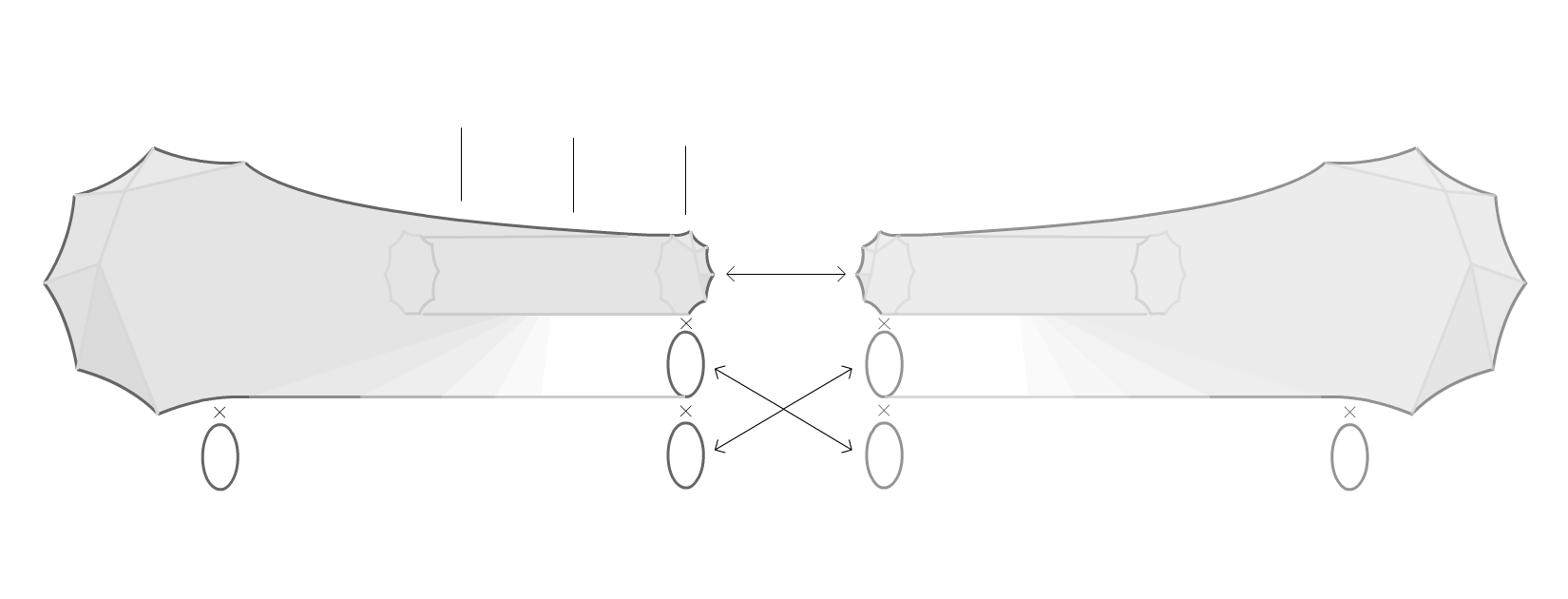}};
\node at (-6.25,0.225) {$X_+$};
\node at (6.3,0.22) {$X_-$};
\node at (-1.15,0.225) {$S_+$};
\node at (1.2,0.22) {$S_-$};
\node at (-1.48,-0.65) {$\mathbb{S}^1_\theta$};
\node at (-1.48,-1.65) {$\mathbb{S}^1_\xi$};
\node at (-6.31,-1.65) {$\mathbb{S}^1_\xi$};
\node at (-6.25,-2.5) {$\underbrace{\qquad\qquad\qquad}$};
\node at (-2.5,-2.5) {$\underbrace{\qquad\qquad\qquad\quad}$};
\node at (6.25,-2.5) {$\underbrace{\qquad\qquad\qquad}$};
\node at (2.5,-2.5) {$\underbrace{\qquad\qquad\qquad\quad}$};
\node at (-6.25,-3) {Region I\textsuperscript{$+$}};
\node at (-2.5,-3) {Region II\textsuperscript{$+$}};
\node at (2.5,-3) {Region II\textsuperscript{$-$}};
\node at (6.25,-3) {Region I\textsuperscript{$-$}};
\node at (0,0.5) {$\mathfrak{r}$};
\node at (-3.3,2) {\tiny $T-1$};
\node at (-2.2,1.85) {\tiny $T$};
\node at (-1,1.8) {\tiny $T+1$};
\end{tikzpicture}
\caption{The twisted connected sum construction of $G_2$ holonomy manifolds}
\label{fig:TCSgeometry}
\end{center}
\end{figure}

First, we need a pair $X_\pm$ of asymptotically cylindrical Calabi--Yau $3$-folds (c.f.\ figure~\ref{fig:TCSgeometry}). A complete Calabi--Yau $n$-fold $(X,g,\omega,\Omega)$ is called \textit{asymptotically (exponentially) cylindrical} (ACyl) if and only if it contains a compact $2n$-dimensional set whose complement is diffeomorphic to a Calabi--Yau half-cylinder $X_\infty=\mathbb{R}^+\times C^{2n-1}$, and if this diffeomorphism identifies sufficiently fast (along $\mathbb{R}^+$) the K\"ahler and holomorphic $n$-forms of $X$ with the corresponding forms $\omega_\infty$ and $\Omega_\infty$ of the half-cylinder. In the TCS case, the cross-section of $X_\infty$ is $C^{5}=S \times \mathbb{S}^1$, where $(S,g_S,\omega_S,\Omega_S)$ is a K3 surface. Parametrising $\mathbb{R}^+$ with $t$ and $\mathbb{S}^1$ with $\theta$, we symbolise the statement about the structure tensors of $X$ as asymptotics:
\begin{align}
g &~ \overset{t\rightarrow\infty}{\widesim} ~ g_\infty = \dd t^2 + \dd\theta^2 + g_S\,, \label{eq:ginfty} \\
\omega &~ \overset{t\rightarrow\infty}{\widesim} ~ \omega_\infty = \dd t \wedge \dd\theta + \omega_S \,, \label{eq:omegainfty} \\
\Omega &~ \overset{t\rightarrow\infty}{\widesim} ~ \Omega_\infty = \left( \dd\theta - i \dd t \right) \wedge \Omega_S \,. \label{eq:Omegainfty}
\end{align}
We will not need much more mathematical rigour for our purposes. It suffices to picture a distinguished real direction in $X$ along which the manifold asymptotes to a translation-invariant Calabi--Yau $X_\infty$ with structure forms as given above. As explained, we need two such ACyl Calabi--Yaus $X_\pm$. We will add $\pm$ to all their associated quantities to distinguish them.

We note that the ambient Calabi--Yau structure induces a particular choice of K\"ahler form $\omega_S$ and holomorphic top form $\Omega_S$ on $S$. Writing $\omega_S = \omega^I$ and $\Omega_S = \omega^J+i\omega^K$, makes the hyper-K\"ahler structure of the K3 surface more manifest. Here the hyper-K\"ahler triplet of closed forms $(\omega^I, \omega^J, \omega^K)$ satisfies by definition the relations 
\begin{align} \label{eq:hyperkahlertriplet}
(\omega^I)^2 = (\omega^J)^2 = (\omega^K)^2 \,, \qquad\qquad
\omega^I\wedge\omega^J = \omega^J\wedge\omega^K = \omega^K\wedge\omega^I = 0 \,.
\end{align}

Now for any Calabi--Yau 3-fold $(X,g,\omega,\Omega)$, it is always possible to equip $\mathcal{M}=X\times \mathbb{S}^1$, with a torsion-free $G_2$-structure. If $\xi$ is the coordinate along the $\mathbb{S}^1$, the associative and co-associative forms on the product are defined respectively as
\begin{equation}
\phi = \dd\xi \wedge \omega + \text{Re}(\Omega)\,, \qquad
\psi = \frac{1}{2}\omega^2 - \dd\xi \wedge \text{Im}(\Omega)\,. \label{eq:M=XxS1--phipsi}
\end{equation}
These are mutually Hodge dual with respect to the product metric
\begin{equation}
g_\mathcal{M} = g + d\xi^2 \,. \label{eq:M=XxS1--g}
\end{equation}
These formul\ae \, characterise a torsion-free $G_2$-structure.  Equivalently, the holonomy of $g_\mathcal{M}$ is contained in $G_2$; here this is a proper inclusion because $\text{Hol}(g_\mathcal{M})=SU(3)$ is only a subgroup of $G_2$. The procedure can be applied equally to $X_\pm$, leading to $(\mathcal{M}_{\pm},g_{\pm},\phi_{\pm},\psi_{\pm})$, and to their asymptotic models $X_{\infty\pm}$, leading to $(\mathcal{M}_{\infty\pm},g_{\infty\pm},\phi_{\infty\pm},\psi_{\infty\pm})$.

The next step is essentially to form the connected sum $\mathcal{M}_+ \,\#\, \mathcal{M}_-$ of the two open $7$-dimensional manifolds by ``pasting'' the asymptotic models. However, this cannot be done in the most naive way because this would lead to a manifold with infinite fundamental group. It is an important fact in $G_2$ geometry that a compact manifold $\mathcal{M}$ with torsion-free $G_2$-structure has holonomy exactly $G_2$ if and only if $\pi_1(M)$ is finite. This is where the ``twist'' plays a useful role. To describe this, we first assume the existence of a diffeomorphism of K3 surfaces\footnote{It is a non-trivial requirement about the ACyl manifolds that they should be compatible in this sense.}
\begin{equation}
\mathfrak{r}\, : S_+ \longrightarrow S_-
\end{equation}
which is an isometry,\footnote{Opening a parenthesis here, the prescription for the behaviour of the B-field
\begin{equation}
\mathfrak{r}^* \, b_{S-} = b_{S+}\,
\end{equation}
given in \cite{Braun:2017csz} to fix the proposal of \cite{Braun:2017ryx}, appears like the correct one from our worldsheet / CFT considerations. Indeed, one would expect automorphisms of the $(1,0)$ nonlinear sigma-model to act on $g$ and $b$ identically, because they appear on the same footing. The map $\mathfrak{r}$ will be interpreted as a semi-classical automorphism of the worldsheet theory later on.}
\begin{equation}
\mathfrak{r}^* \, g_{S-} = g_{S+}\,
\end{equation}
and which satisfies
\begin{equation}
\mathfrak{r}^* \, \text{Im}(\Omega_{S-}) = -\text{Im}(\Omega_{S+})\,,\qquad
\mathfrak{r}^* \, \text{Re}(\Omega_{S-}) = \omega_{S+}\,,\qquad
\mathfrak{r}^* \, \omega_{S-} = \text{Re}(\Omega_{S+})\,.
\end{equation}

Such a map is called a \textit{hyper-K\"ahler matching}. We use it to fuse $\mathcal{M}_+$ and $\mathcal{M}_-$ together. More precisely, let $T\in \mathbb{R}^+$ and introduce a boundary at $t=T+1$ to define the manifolds $\mathcal{M}_\pm(T)$ with boundary $\partial \, \mathcal{M}_\pm(T) \simeq S_\pm \times \mathbb{S}^1_\pm \times \mathbb{S}^1_\pm$. ($T$ is eventually taken large enough for existence theorems to apply.) Let $I=[T,T+1]\subset \mathbb{R}^+$. Next, introduce a diffeomorphism
\begin{equation} \label{eq:GeomPastingMap}
F_T \quad : \quad \begin{matrix}
I\times S_+ \times \mathbb{S}^1_+ \times \mathbb{S}^1_+ & \longrightarrow & I\times S_- \times \mathbb{S}^1_- \times \mathbb{S}^1_- \\ 
(~t~, ~z~, ~\theta~, ~\xi~) & \longmapsto & (~2T+1-t~, ~\mathfrak{r}(z)~, ~\xi~, ~\theta~)\,.
\end{matrix}
\end{equation}
It is a simple exercise to show that the associative and co-associative forms are compatible under this map,
\begin{equation}
(F_T)^* \, \phi_{\infty-} = \phi_{\infty+}\,, \qquad (F_T)^* \, \psi_{\infty-} = \psi_{\infty+} \,.
\end{equation}
Since the $G_2$ metric can be derived from any of these, this implies in particular that $F_T$ is an isometry.

The pasting is finished as follows. Endow $\mathcal{M}_\pm(T)$ with the $G_2$-structure interpolating from $(g_\pm,\phi_\pm,\psi_\pm)$ at $t=T-1$ to $(g_{\infty\pm},\phi_{\infty\pm},\psi_{\infty\pm})$ at $t=T$. Next, on $I=[T,T+1]$, use $F_T$ to establish the local isomorphism $\mathcal{M}_{\infty +} \simeq \mathcal{M}_{\infty -}$. This yields a compact $G_2$-structure manifold $\mathcal{M}=\mathcal{M}_+ \,\#_{\mathfrak{r}}\, \mathcal{M}_-$ with a neck-like region of length $\sim 2T$. There remains non-vanishing torsion concentrated in $t\in[T-1,T]$ on both sides. Nevertheless there are theorems that guarantee, for sufficiently large $T$, the existence of a torsion-free and Ricci-flat perturbation of the $G_2$-structure $\phi$ of $\mathcal{M}$ within the same cohomology class (c.f.\ \cite{Joyce2000} theorem 11.6.1, \cite{MR2024648} theorem 5.34, and \cite{Corti:2012kd} theorem 3.12).

Note that the larger the parameter $T$, the lesser the perturbation needed to reach vanishing torsion. A TCS with $T$ very large, i.e.\ with a highly stretched ``neck'', is said to be in a \textit{Kovalev limit}. This is often a useful limit to work with in practical applications. As is manifest in section~\ref{sec:Realisations}, our algebraic construction however has no notion of scale $T$, so a priori no Kovalev limit. This suggests that it remains adequate even in the limit of a short neck, i.e.\ in the phase of the conformal manifold where geometric control starts to disappear.

\subsection{Shatashvili--Vafa $G_2$ superconformal algebra} \label{sec:SCAandG2}

We encounter a few different examples of superconformal algebras in the next section, such as free fermions and bosons, and Odake algebras. By and large, our focus is however on the so-called Shatashvili--Vafa superconformal algebra associated to $G_2$, which we now briefly describe. For clarifications about superconformal algebras themselves, we refer to appendix~\ref{app:algebras}.

As a higher spin $\mathcal{W}$-algebra, the $G_2$ SCA was first noticed in \cite{Blumenhagen:1991nm, Blumenhagen:1992vr}. It is part of a $2$-parameter family \cite{Noyvert:2002mc} denoted by $\mathcal{SW}^{\mathcal{N}=1}(\tfrac{3}{2},\tfrac{3}{2},2)$ in the nomenclature of \cite{Bouwknegt:1992wg}. Apart from an $\mathcal{N}=1$ superconformal multiplet, generating Virasoro with $c=\tfrac{21}{2}$, its generators are denoted $(\Phi, K)$ and $(X, M)$, where we have formed $\mathcal{N}=1$ pairs. The conformal weights of $\Phi$, $K$, $X$, and $M$ are respectively $\tfrac{3}{2}$, $2$, $2$, and $\tfrac{5}{2}$. All the explicit OPE relations are provided in appendix~\ref{app:algebras}. An intriguing peculiarity of this algebra is that it contains another $\mathcal{N}=1$ Virasoro besides $(G,T)$. The pair $(\Phi,X)$ realises the $c =\tfrac{7}{10}$ SCA which is the only minimal model for both $\mathcal{N}=0$ and $\mathcal{N}=1$ Virasoro and which describes the fixed locus of the tri-critical Ising model.

An important contribution of Shatashvili and Vafa in \cite{Shatashvili:1994zw} is the realisation of this algebra $\left\langle T,G,\Phi,K,X,M \right\rangle$ using free fermions $\psi^i$ and bosons $j^i$, $i\in\{1,2,\ldots,7\}$. The most striking by-product of this realisation is that it makes manifest the relationship with $G_2$-structure manifolds. Indeed, the formul\ae \, for $\Phi$ in terms of the fermions $\psi^i$ is identical to the definition of the canonical $G_2$ $3$-form on $\mathbb{R}^7$:
\begin{equation}
\phi_0=\dd x^{125} + \dd x^{136} + \dd x^{147} - \dd x^{237} + \dd x^{246} - \dd x^{345} + \dd x^{567} \,,
\end{equation}
provided we identify $\psi^i$ with $\dd x^i$ and normal ordering with the wedge product. Similarly, the generator $X$ can essentially be thought of as a quantum version of the co-associative $4$-form $\psi$.\footnote{\label{foot:X} Strictly speaking, there are quantum corrections in the case of the $4$-form. In both realisations of \cite{Shatashvili:1994zw} and \cite{Figueroa-OFarrill:1996tnk} (described later), the $4$-form operator $\Psi$, which is the classical current \cite{Howe:1991ic,delaOssa:2018azc}, differs from the quantum generator $X$ by the fermionic energy-momentum tensor associated to the flat directions in the geometry.} This kind of correspondence between geometric and algebraic quantities becomes useful later when we discuss the TCS realisation of the $G_2$ SCA.

There is an aspect of the $G_2$ SCA that cannot be captured by its free field realisation. Abstractly the OPE relations that define it are only associative modulo the ideal generated by the null field \cite{Figueroa-OFarrill:1996tnk}
\begin{equation}\label{eq:NullFieldinG2SCA}
N=4\no{GX}-2\no{\Phi K}-4M'-G'' \,,
\end{equation}
where colons represent normal ordering.
 
\section{Conformal algebras and twisted connected sums} \label{sec:Realisations}

In this section we explain how to realise the Shatashvili--Vafa $G_2$ algebra associated to twisted connected sum geometries. Achieving this involves two main steps: 
\begin{enumerate}
\item Find realisations of the $G_2$ SCA associated to distinctive local open subsets $U_\alpha$ in the TCS geometry;
\item Match these local realisations on overlaps $U_\alpha\cap U_\beta$ in the TCS geometry by using invariances.
\end{enumerate}
This strategy may seem unusual in a number of ways because it mixes intuition from differential geometry and conformal field theory. We address this here, starting with point 1.

One often regards SCAs as providing an exact quantum description of chiral symmetries of a conformal field theory used as internal sector in a string compactification. This is true whether or not the internal theory admits a geometric interpretation as the infra-red fixed point of a non-linear sigma model. In particular, this abstract framework should allow a description of non-geometric phases of the internal theory, where the ``local open subsets'' of point 1 are non-existent.

We suggest to resolve this puzzle by assuming that we are working in an intermediate phase of the conformal manifold between the perturbative non-linear sigma model, whose target manifold is approximately\footnote{A $G_2$ holonomy target does not lead to a scale-invariant model due to corrections in $\alpha'$. However, it was argued not long ago \cite{Becker:2014rea} that such corrections only mildly perturb away from $G_2$ holonomy.} a TCS, and the full-fledged quantum regime. In this intermediate region, there should be remnants of a geometric intuition about the would-be target space, and thus it makes sense to refer to local open subsets. However, we are seeking to venture into the stringy regime and the description is not in terms of a non-linear sigma model, but rather in terms of ``local'' SCAs. This understanding is at least useful to bear in mind for now.

There are two qualitatively different regions in each of the two parts $\mathcal{M}_\pm$ of the TCS construction. One is approximated by the direct product of an ACyl Calabi--Yau $3$-fold with an $\mathbb{S}^1$ (type $\text{I}^\pm$ in figure~\ref{fig:TCSgeometry}) and the other one is the asymptotic region $\mathcal{M}_{\infty\pm}=\mathbb{R}^+ \times S_\pm \times \mathbb{S}^1 \times \mathbb{S}^1$ (type $\text{I}^\pm$). We will show below that each region admits a distinguished local realisation of the general $G_2$ algebra highly reminiscent of the associated geometric structure.

Step 2 above also deserves more explanations. We take again a semi-classical point of view and borrow from the non-linear sigma model description. With sigma models, it is often customary to designate a \textit{compact} manifold $\mathcal{M}$ as target space. In the strictest of senses this cannot be done explicitly (i.e.\ with a Lagrangian) unless the target space has particular symmetries, such as for group manifolds. For a generic target space, a particular open patch $U_\alpha\subset \mathcal{M}$ and local coordinates must indeed be chosen before a Lagrangian $\mathcal{L}_\alpha$ can be formulated. 

This technical limitation is quite inconsequential in practise because all the local Lagrangians $\mathcal{L}_\alpha$ are sufficiently consistent with one another for a well-defined theory $\mathcal{L}=\bigcup_{U_\alpha\in \mathcal{M}}\mathcal{L}_\alpha$ to exist for the whole of $\mathcal{M}$. The key feature of $\mathcal{L}_\alpha$ allowing this global interpretation is the existence of symmetries (or more general formal invariances involving variations of the couplings of the sigma model). These invariances of $\mathcal{L}_\alpha$ admit target space interpretations, such as for example diffeomorphisms, isometries, gauge invariances, etc. To insist, it is these invariances of $\mathcal{L}_\alpha$ that inform us on how to interpret globally the theory.\footnote{A parallel can be drawn here with the discussion in \cite{QFT&Strings2} (lectures 17 and 18) motivating the introduction of a Seiberg--Witten curve to describe the low-energy dynamics of $\mathcal{N}=2$ gauge theories in four dimensions. In that case the sigma model is an $\mathcal{N}=2$ $U(1)$ gauge theory and the $SL(2,\mathbb{Z})$ invariance of the field-dependent coupling $\tau_{\text{IR}}$ suggests an elliptic fibration over the Coulomb branch.} The proposal of point 2 above is that invariances of the local realisations of the $G_2$ SCA should similarly allow us to construct a globally consistent realisation for the whole TCS geometry. We will make this fully precise in section~\ref{sec:Compatibility}. We must first describe the realisations corresponding to the two types of regions.

\subsection{Type $\text{I}^\pm$: $X_\pm \times \mathbb{S}^1$} \label{sec:TypeI}

The superconformal algebra $\text{Od}^n$ associated to Calabi--Yau $n$-folds was first described by Odake \cite{Odake:1988bh}. It is an extension of $\mathcal{N}=2$ Virasoro by the spectral flow operator, which we denote by $\Omega=A+iB$. This field is primary with weight $n/2$ relatively to the energy-momentum tensor $T$ and it is related to the holomorphic $n$-form on the manifold \cite{Howe:1991ic}. Its superpartner is $\Upsilon=(C+iD)/\sqrt{2}$. The remaining $\mathcal{N}=2$ generators are the $U(1)_R$ current $J$ (corresponding to the K\"ahler form) and the supercurrents $G=G^0$ and $G^3$. Similarly to the $G_2$ SCA, associativity of the OPEs only holds in the case $n=3$ modulo an ideal. Here this ideal is generated by the fields
\begin{equation} \label{eq:N1N2}
N^{1}=A'-\no{JB}\,,\qquad N^{2}=B'+\no{JA}\,.
\end{equation}
Our normalisations can be inferred from the explicit OPEs provided in appendix~\ref{app:algebras}.

Associated to the $\mathbb{S}^1$ factor, we can realise $\mathcal{N}=1$ Virasoro in the usual way using free fields. We parametrise the $\mathbb{S}^1$ by the free boson $\xi$ and define the associated $U(1)$ current as $j_\xi=i\xi'$. Its superpartner is a Majorana--Weyl fermion denoted by $\psi_{\xi}$. The $\mathcal{N}=1$ generators are given explicitly by
\begin{align}
T_{\xi}&=\tfrac{1}{2}\no{j_\xi j_\xi}+\tfrac{1}{2}\no{\psi_{\xi}' \psi_{\xi}}\,, \label{eq:Tflat}\\
G_{\xi}&=\no{j_\xi \psi_{\xi}}\,. \label{eq:Gflat}
\end{align}
The free field singular OPEs are
\begin{equation} \label{eq:freeOPEs}
\wick{\c j_\xi(z) \c j_\xi(w)} = \frac{1}{(z-w)^2}\,,\qquad\qquad
\wick{\c \psi_{\xi}(z) \c \psi_{\xi}(w)} = \frac{1}{z-w}\,.
\end{equation}

These two algebras --- free fields and $\text{Od}^{n=3}$ --- can be combined to provide a realisation of the $G_2$ SCA. It will be regarded as the local form taken by the $G_2$ SCA in the geometric regions $\text{I}^\pm$ in figure~\ref{fig:TCSgeometry}. As first shown in \cite{Figueroa-OFarrill:1996tnk}, the following generators\footnote{Whenever confusion between generators of different algebras might occur, we use indices to distinguish them. Here $T_\mathcal{M}$ and $T$ are respectively energy-momentum tensors for the $G_2$ and Odake $n=3$ SCAs.} 
\begin{align}
T&_\mathcal{M}=T+T_{\xi} \label{eq:T7} \\
G&_\mathcal{M}=G+G_{\xi} \label{eq:G7} \\
\Phi&=\no{J \psi_{\xi}}+A \label{eq:Phi7} \\
X&=\no{B \psi_{\xi}}+\tfrac{1}{2}\no{JJ}-\tfrac{1}{2}\no{\psi_{\xi}' \psi_{\xi}} \label{eq:X7} \\
K&=C+\no{J j_\xi}+\no{G^3 \psi_{\xi}} \\
M&=\no{D \psi_{\xi}}-\no{B j_\xi}+\tfrac{1}{2}\no{j_\xi \psi'_{\xi}}-\tfrac{1}{2}\no{j'_\xi \psi_{\xi}}+\no{JG^3}-\tfrac{1}{2}G' \label{eq:M7}
\end{align}
satisfy the $G_2$ OPE relations, up to the ideal generated by $N^{1}$ and $N^{2}$. We summarise this key statement as
\begin{equation} \label{eq:InclusionI}
G_2\,\text{SCA} \overset{\text{I}}{\longhookrightarrow} \left(\text{Od}^{n=3}\times \text{Free}\right) ~~ \Big/ ~ \langle N^1 , N^2\rangle \,.
\end{equation}
Moreover, the null field $N$ in the $G_2$ algebra (c.f.\ \eqref{eq:NullFieldinG2SCA}) belongs to the $\text{Od}^{n=3}$ ideal:
\begin{equation}
\langle N \rangle \subset \langle N^1 , N^2\rangle\,.
\end{equation}

In the work of \cite{Figueroa-OFarrill:1996tnk}, this realisation was motivated by the construction of $G_2$ holonomy manifolds as desingularisations of $\mathbb{Z}_2$ quotients of Calabi--Yau $3$-folds times a circle \cite{MR1424428, Joyce2000}. It is a property of the realisation that it is left invariant by a $\mathbb{Z}_2$ map corresponding to the quotient, and hence it does descent to the quotient as stated in \cite{Figueroa-OFarrill:1996tnk}. We have verified however that the OPEs themselves do not rely on the $\mathbb{Z}_2$  identification and therefore hold for $X_\pm \times \mathbb{S}^1$ even though the holonomy of this space is only contained in $G_2$.

We should point out a minor discrepancy between our generator $M$ in \eqref{eq:M7} and the one given in \cite{Figueroa-OFarrill:1996tnk}. We believe this is due to a sign error. 

Let us also appreciate the similarity between the field theoretic definitions \eqref{eq:T7} and \eqref{eq:G7} of $T_\mathcal{M}$ and $G_\mathcal{M}$ with the direct product metric \eqref{eq:M=XxS1--g} on $\mathcal{M_\pm}$. The same comment applies to $\phi$ and $\psi$ in \eqref{eq:M=XxS1--phipsi} which are essentially identical to $\Phi$ and $X$, up to the identifications
\begin{equation}
\psi_\xi \leftrightarrow \dd \xi \,, \qquad \no{~} \leftrightarrow \wedge \,, \qquad (J,A,B) \leftrightarrow (\omega , \text{Re}(\Omega) , \text{Im}(\Omega)) \,. \label{eq:GeometryVsSCA}
\end{equation}
The only difference is the quantum correction in the case of $X$ already mentionned in footnote~\ref{foot:X}.

The embedding of the $G_2$ algebra described by \eqref{eq:InclusionI}--\eqref{eq:M7} is not unique. This is due to the automorphism $A+iB\longmapsto e^{iu}(A+iB)$, $u\in \mathbb{R}$, (with a similar transformation for $\Upsilon$) of the Odake algebra, which can be used to rotate the $\{A,B\}$ basis into itself. We will discuss automorphisms more fully in section~\ref{sec:Automorphisms}.

\subsection{Type $\text{II}^\pm$: $\mathbb{R}^+ \times S_\pm \times \mathbb{S}^1 \times \mathbb{S}^1$} \label{sec:TypeII}

In this region, the geometric intuition calls first for a realisation of a copy of $\text{Od}^{n=3}$ associated to the translation-invariant Calabi--Yaus $X_{\infty\pm}$ in terms of the algebras corresponding to  $\mathbb{R}^+$, $S_\pm$, and $\mathbb{S}^1$. Then, by the results of section~\ref{sec:TypeI}, we will be able to define a realisation of the $G_2$ SCA associated to $\mathcal{M}_{\infty\pm}=X_{\infty\pm}\times \mathbb{S}^1$. We start by examining how to realise the $\text{Od}^{n=3}$ for $X_{\infty\pm}$.

Each flat factor in the geometry leads again to free fields and Virasoro generators using \eqref{eq:Tflat}--\eqref{eq:freeOPEs}. We will keep denoting by $\xi$ the coordinate on the external $\mathbb{S}^1$ and we reuse the coordinates $t$, $\theta$ from section~\ref{sec:TCSgeometry} for $\mathbb{R}^+$ and the internal $\mathbb{S}^1$ respectively.

To the K3 surface, a Calabi--Yau $2$-fold, corresponds the $n=2$ Odake algebra. In this case, the generator $\Omega=A+iB$ extending $\mathcal{N}=2$ Virasoro forms with $J$ an $SU(2)$ Kac-Moody current algebra at level $k=1$. The superpartner of $\Omega$ is a complex supersymmetry current. The $n=2$ Odake SCA is isomorphic to the so-called \textit{little} $\mathcal{N}=4$ Virasoro SCA \cite{Odake:1988bh,Ali:1999ut,Ademollo:1976wv,Ademollo:1975an}.

Considering now the asymptotic formul\ae~\eqref{eq:ginfty}--\eqref{eq:Omegainfty} for $g_\infty$, $\omega_\infty$ and $\Omega_\infty$ in region II, and taking stock of the correspondences between geometry and SCAs stated in \eqref{eq:GeometryVsSCA}, we are naturally led to some ans\"atze for generators of $\text{Od}^{n=3}$. We take
\begin{align}
T&=T_t+T_S+T_{\theta} \,,\label{eq:T3}\\
G&=G_t+G_S+G_{\theta} \,,\label{eq:G3}\\
J&=\no{\psi_t \psi_{\theta}}+J_S \,,\label{eq:J3}\\
A+iB&=\no{(\psi_{\theta}-i\psi_t) (A_S+iB_S)} \,,\label{eq:A3+iB3}
\end{align}
where the subscript $S$ identifies the generators of $\text{Od}^{n=2}$. The remaining generators can easily be derived by taking OPEs with $G$ in \eqref{eq:G3}. We find
\begin{align}
G^3&=G^3_S+\no{\psi_{\theta}j_t}-\no{j_\theta \psi_t} \,, \label{eq:G33}\\
C&=\no{j_t B_S}+\no{j_\theta A_S}-\no{\psi_tD_S}-\no{\psi_\theta C_S} \,, \label{eq:C3}\\
D&=-\no{j_t A_S}+\no{j_\theta B_S}+\no{\psi_tC_S}-\no{\psi_\theta D_S} \,. \label{eq:D3}
\end{align}

It can then be checked that these generators indeed give rise to the Odake $n=3$ algebra. Some of the OPEs hold exactly, while others only hold modulo the ideal generated by the fields in \eqref{eq:N1N2}, themselves written in terms of  $\text{Od}^{n=2}$ and free generators:
\begin{align}
N^1 &= \no{\psi_t\no{J_SA_S}}+\no{\psi_tB_S'}-\no{\psi_\theta\no{J_SB_S}}+\no{\psi_\theta A_S'}\,, \label{eq:N1II} \\
N^2 &= \no{\psi_t\no{J_SB_S}}-\no{\psi_tA_S'}+\no{\psi_\theta\no{J_SA_S}}+\no{\psi_\theta B_S'}\,. \label{eq:N2II}
\end{align}
We write this result as
\begin{equation} \label{eq:InclusionIICY}
\text{Od}^{n=3} \overset{\text{II}_X}{\longhookrightarrow} \left(\text{Od}^{n=2} \times (\text{Free})^2\right) ~~ \Big/ ~ \langle N^1 , N^2\rangle \,.
\end{equation}

As an example of the implications of the ideal, let us examine the OPE of $A$ with itself. The expected result in the Odake $n=3$ algebra is
\begin{equation}
\wick{ \c A(z) \c A(w)} = -\frac{4}{(z-w)^3}-\frac{2\no{JJ}(w)}{z-w}\,.
\end{equation}
Computing directly with \eqref{eq:A3+iB3}, the result at order $-1$ differs from the expectation by
\begin{equation} \label{eq:AAmismatch}
\no{A_SA_S}+\no{B_SB_S}-2\no{J_SJ_S}\,.
\end{equation}

That this should vanish nicely reflects the hyper-k\"ahler structure of K3 surfaces. The geometric interpretation of $J_S$, $A_S$ and $B_S$ is as a triplet $(\omega^I,\omega^J,\omega^K)$ of $(1,1)$-forms related by the condition~\eqref{eq:hyperkahlertriplet}:
\begin{equation}
(\omega^I)^2 = (\omega^J)^2 = (\omega^K)^2 \,.
\end{equation}
We can actually derive this hyperk\"ahler condition directly from the structure of the conformal algebra owing to the identification $\no{~}\leftrightarrow \wedge$. Since we should consider the algebra only modulo the ideal $\langle N^1, N^2\rangle$ generated by the fields in \eqref{eq:N1N2}, any fields appearing in OPEs involving $N^1$ or $N^2$ can be taken to vanish. After some work, we find
\begin{align}
\frac{1}{2} \, \wick{\no{A_S \c \psi_\theta}(z) \; \c N^1(w)}&=\frac{\no{J_SJ_S}(w)-\no{B_SB_S}(w)}{(z-w)^2}+\ldots \\
\frac{1}{2} \, \wick{\no{B_S \c \psi_\theta}(z) \; \c N^2(w)}&=\frac{\no{J_SJ_S}(w)-\no{A_SA_S}(w)}{(z-w)^2}+\ldots
\end{align}
proving in particular that \eqref{eq:AAmismatch} belongs to the ideal.

With the Odake algebra associated to $X_{\infty\pm}$ now established, we can proceed as in section~\ref{sec:TypeI} and define $G_2$ SCA generators by substituting the $n=3$ Odake generators \eqref{eq:T3}--\eqref{eq:D3} in \eqref{eq:T7}--\eqref{eq:M7}. For instance, we get
\begin{align} \label{eq:Phi7eg}
\Phi=\no{\no{\psi_t\psi_{\theta}}\psi_{\xi}}+\no{J_S\psi_{\xi}}+\no{\psi_{\theta}A_S}+\no{\psi_t  B_S} \,.
\end{align}
Again the $G_2$ OPEs are realised up to the ideal generated by $N^1$ and $N^2$. This yields the realisation
\begin{equation} \label{eq:InclusionII}
G_2\,\text{SCA} \overset{\text{II}}{\longhookrightarrow} \left(\text{Od}^{n=2} \times (\text{Free})^3\right) ~~ \Big/ ~ \langle N^1 , N^2\rangle \,.
\end{equation}

These embeddings are not unique due to the automorphisms of $\text{Od}^{n=2}$ and $\text{Od}^{n=3}$ mentioned at the end of section~\ref{sec:TypeI}. There is in fact a $2$-parameter family of possible realisations of the $G_2$ algebra in region II and we have selected one for definitiveness.

\subsection{Compatibility at junctions} \label{sec:Compatibility}

We distinguish three regions in the TCS geometry where the local patches supporting a known realisation of the $G_2$ SCA overlap: $\text{I}^+\cap \text{II}^+$, $\text{II}^+\cap \text{II}^-$, and $\text{I}^-\cap \text{II}^-$. We now wish to establish transition maps on such overlaps. Let us start with $\text{I}^+\cap\text{II}^+$. ($\text{I}^-\cap\text{II}^-$ is treated identically.)

It should be rather clear from our exposition in section~\ref{sec:Realisations} that the realisations on $\text{I}^\pm\cap \text{II}^\pm$ are compatible. There is a clear map between the abstract generators $\{T,G,G^3,J,A,B,C,D\}$ in region $\text{I}^\pm$ and the explicit generators which they asymptote to in region $\text{II}^\pm$ as $t\rightarrow\infty$, c.f.\ \eqref{eq:T3}--\eqref{eq:D3}. Under this correspondence, which is linear and bijective, the OPE relations of $\text{Od}^{n=3}$, and thus of the $G_2$ SCA, are invariant. Such a map preserving the algebraic structure defined by the OPEs will henceforth be called an \textit{automorphism} of the SCA.

It remains to ascertain algebraic compatibility over the patch in the TCS geometry where the two parts merge into one another. On each side ($+/-$) of the joint, the $G_2$ algebra is realised in the same way (type $\text{II}$). We use $\pm$ labels to distinguish the generators. Geometrically, the isomorphism $F_T$ in \eqref{eq:GeomPastingMap} allows to assemble the two sides. Algebraically, this should correspond again to an automorphism of the $G_2$ algebra. We thus consider the map
\begin{equation}
F:\begin{cases}
(j_{t+},\psi_{t+})           &\longmapsto \quad (-j_{t-},-\psi_{t-})         \\
\\
(T_{S+},G_{S+})              &\longmapsto \quad (T_{S-},G_{S-})              \\
\qquad\quad J_{S+}           &\longmapsto \quad A_{S-}                       \\
\qquad\quad A_{S+}           &\longmapsto \quad J_{S-}                       \\
\qquad\quad B_{S+}           &\longmapsto \quad -B_{S-}                      \\
\qquad\quad G^3_{S+}         &\longmapsto \quad C_{S-}                       \\
\qquad\quad C_{S+}           &\longmapsto \quad G^3_{S-}                     \\
\qquad\quad D_{S+}           &\longmapsto \quad -D_{S-}                      \\
\\
(j_{\xi+},\psi_{\xi+})       &\longmapsto \quad (j_{\theta-},\psi_{\theta-}) \\
(j_{\theta+},\psi_{\theta+}) &\longmapsto \quad (j_{\xi-},\psi_{\xi-}) 
\end{cases} \label{eq:PastingAuto}
\end{equation}

It is easy to see that $T_{\mathcal{M}+}\longmapsto T_{\mathcal{M}-}$ and $G_{\mathcal{M}+}\longmapsto G_{\mathcal{M}-}$ under these transformations (c.f.\ \eqref{eq:T7}, \eqref{eq:G7}, and \eqref{eq:T3}--\eqref{eq:D3}). We can also straightforwardly verify that the generator $\Phi$ in~\eqref{eq:Phi7eg} is preserved:
\begin{equation}
\Phi_+ \longmapsto \left(-\no{\no{\psi_t\psi_{\xi}}\psi_{\theta}}+\no{A_S\psi_{\theta}}+\no{\psi_{\xi}J_S}+\no{\psi_t B_S}\right)_- = \Phi_-\,,
\end{equation}
where the last equality follows from associativity of normal ordering and skew-commutativity for free fermions. Since $T_\mathcal{M}$, $G_\mathcal{M}$ and $\Phi$ can be used to define the remaining generators ($X$, $K$, $M$), it may seem clear from here that $F$ defines an automorphism by sending all generators to themselves. However, some more work is needed to prove this statement\footnote{We remove the $\pm$ to avoid cluttering the proof.} since the fields $N^1$, $N^2$ (c.f.\ \eqref{eq:N1II}--\eqref{eq:N2II}) get acted upon non-trivially by the map $F$:
\begin{align}
N^1 &\longmapsto \tilde{N}^1 \equiv -\no{\psi_t\no{A_SJ_S}}+\no{\psi_tB_S'}+\no{\psi_\xi\no{A_SB_S}}+\no{\psi_\xi J_S'}\,, \\
N^2 &\longmapsto \tilde{N}^2 \equiv \no{\psi_t\no{A_SB_S}}+\no{\psi_tJ_S'}+\no{\psi_\xi\no{A_SJ_S}}-\no{\psi_\xi B_S'}\,.
\end{align}
Recall that modding out by these fields is essential for associativity, so the status of associativity is a priori unclear.

We can show however that the original ideal $\langle N^1, N^2\rangle$ generated by $N^1$ and $N^2$ is in fact isomorphic to the ideal $\langle \tilde{N}^1, \tilde{N}^2\rangle$. It is also straightforward to see explicitly that all the remaining generators \{$X$, $K$, $M$\} are preserved by the map \eqref{eq:PastingAuto}, which guarantees that $F$ is indeed an automorphism of the $G_2$ SCA.

To prove the equivalence of the ideals, we notice first from \eqref{eq:N1II} and \eqref{eq:N2II} that $N^1$ and $N^2$ are contained in the ideal generated by
\begin{equation}
N_S^1 = A_S'-\no{J_SB_S}\,,
\end{equation}
\begin{equation}
N_S^2 = B_S'+\no{J_SA_S}\,.
\end{equation}
Thus $\langle N^1, N^2\rangle \subseteq \langle N_S^1, N_S^2\rangle$. On the other hand, $N_S^1$, $N_S^2$ may be found in $\langle N^1, N^2\rangle$, which proves the other inclusion. Indeed we have
\begin{align}
\wick{ \c \psi_\theta(z) \c N^1(w)}&=\frac{N_S^1(w)}{z-w}\,, \\
\wick{ \c \psi_\theta(z) \c N^2(w)}&=\frac{N_S^2(w)}{z-w}\,.
\end{align}

Hence $\langle N^1, N^2\rangle = \langle N_S^1, N_S^2\rangle$. We can then prove $\langle \tilde{N}^1, \tilde{N}^2\rangle = \langle N_S^1, N_S^2\rangle$ as follows. First we note that $\no{[A_S,J_S]}=2B_S'$ (following from $\text{Od}^{n=2}$). From this we see that $\tilde{N}^1$ and $\tilde{N}^2$ are normal ordered products
\begin{align}
\tilde{N}^1 &= \no{\psi_t N^2_S}+\no{\psi_\theta N^3_S} \\
\tilde{N}^2 &= \no{\psi_t N^3_S}-\no{\psi_\xi N^2_S}
\end{align}
of free fermions with $N^2_S$ and the field
\begin{equation}
N^3_S=J_S'+\no{A_SB_S}\,,
\end{equation}
which itself arises as
\begin{equation}
\frac{1}{2} \, \wick{\no{A_S \c \psi_\theta}(z) \; \c N^2(w)} =\frac{N^3_S(w)}{(z-w)^2}+\ldots
\end{equation}
This shows $\langle \tilde{N}^1, \tilde{N}^2\rangle \subseteq \langle N_S^1, N_S^2\rangle$. We get the remaining inclusion as follows:
\begin{align}
\frac{1}{2} \, \wick{\no{J_S \c \psi_\xi}(z) \; \c {\tilde{N}^2}(w)} &=\frac{N_S^1(w)}{(z-w)^2}+\ldots \\
\wick{ \c \psi_\xi(z) \c {\tilde{N}}^2(w)} &=\frac{N_S^2(w)}{z-w}\,.
\end{align}

Summarising, in spite of the fact that $N^1$ and $N^2$ are not invariant under $F$, the ideal they generate is. We write this as
\begin{equation}
\langle N^1 , N^2 \rangle \subset \frac{\text{Od}^{n=2}\times (\text{Free})^3}{F}  \,.
\end{equation}
Moreover, all the generators of the $G_2$ SCA are invariant, which guarantees that the map acts as the identity automorphism at the level of the $G_2$ SCA. Alternatively, the realisation defined by \eqref{eq:InclusionII} sits in the quotient by $F$:
\begin{equation} \label{eq:resultForIntro}
G_2\,\text{SCA} \longhookrightarrow \frac{\text{Od}^{n=2} \times (\text{Free})^3}{F} ~~ \Big/ ~ \langle N^1 , N^2\rangle \,.
\end{equation}

\section{Automorphisms and $G_2$ mirror symmetry} \label{sec:Automorphisms}

Multiple conformal algebras are delicately incorporated in the TCS realisation of the $G_2$ algebra presented in the last section. Each of these admit automorphisms and some of them are known to be related to T-duality and Calabi--Yau mirror symmetry in type II string theory \cite{Lerche:1989uy}. For example, some of the early hints of the existence of Calabi--Yau mirrors were revealed by applying the automorphism
\begin{equation} \label{eq:Calabi-YauMiSYAut}
J\longmapsto -J\,, \qquad G^3\longmapsto -G^3
\end{equation}
on, say, the right-chiral sector of $\mathcal{N}=(2,2)$ supersymmetric theories (keeping the left-movers invariant).

It is natural to ask if similar automorphisms are associated to mirror symmetry in the $G_2$ case, at least in the examples suggested of this phenomenon. Mirror symmetry for Joyce orbifolds received most of the attention \cite{Acharya:1996fx,Acharya:1997rh,Papadopoulos:1995da,Gaberdiel:2004vx}. In particular in \cite{Gaberdiel:2004vx}, by generalising the concept of discrete torsions \cite{Vafa:1994rv}, the authors were able to interpret all the then-known $G_2$ mirror dualities as the result of an automorphism $\bm{\mathcal{M}}$ (see below) applied on the (right-chiral) $G_2$ SCA. In the examples they covered, this automorphism was generated by T-dualities along the flat directions.

Equipped with our construction from section~\ref{sec:Realisations}, we are now in a position to get a similar conformal field theoretic understanding of mirror symmetry for twisted connected sums \cite{Braun:2017ryx,Braun:2017csz}. To this end, we systematically consider in this section some automorphisms of the algebras discussed above: free fields, $\text{Od}^{n=2}$, $\text{Od}^{n=3}$, and the $G_2$ SCA. Then we address how these automorphisms may be composed together while respecting the compatibility maps established in section~\ref{sec:Compatibility}. This yields two ``mirror'' maps valid for the whole TCS realisation. We find striking similarities with the geometric maps $\mathcal{T}_4$ and $\mathcal{T}_3$ proposed in \cite{Braun:2017ryx,Braun:2017csz} and we connect with the interpretation in \cite{Gaberdiel:2004vx}.

\subsection{Basic automorphisms and their relations} \label{sec:AutomBasic}

We start by listing some simple automorphisms of the algebras used previously which hold regardless of any particular realisation. Almost all of them consist in sign shifts so it is convenient to display them in tables. As an example of how to read the tables, the \textit{parity} automorphism in the first row of the first table translates in
\begin{equation}
\mathbb{\mathbf{P}}_\xi~:~\begin{cases}
j & \longmapsto +j \,, \\
\psi & \longmapsto -\psi \,.
\end{cases}
\end{equation}
All of these automorphisms can be checked easily by considering the explicit OPE relations in appendix~\ref{app:algebras}.

\subsubsection*{Free fields:}

$\qquad$\renewcommand\arraystretch{1.2}
\begin{tabular}{ r | c | c }
 & $j_\xi$ & $\psi_\xi$ \\\hline
Parity $\textbf{P}_\xi$ & $+$ & $-$ \\\hline
T-duality $\textbf{T}_\xi$ & $-$ & $-$
\end{tabular}
\label{tab:Free}

\vspace{10pt}

As the name suggests, when $\xi$ parametrises an $\mathbb{S}^1$, the second map above applied on right-movers is a T-duality.

\vspace{5pt}

\subsubsection*{Odake $n=2$:}

$\qquad$\renewcommand\arraystretch{1.2}
\begin{tabular}{ r | c | c | c | c | c | c | c | c }
 & $T$ & $G$ & $G^3$ & $J$ & $A$ & $B$ & $C$ & $D$ \\\hline
Parity $\textbf{P}_S$ & $+$ & $-$ & $-$ & $+$ & $+$ & $+$ & $-$ & $-$ \\\hline
Mirror symmetry $\textbf{M}_S$ & $+$ & $+$ & $-$ & $-$ & $+$ & $-$ & $+$ & $-$ \\\hline
Phase $\textbf{Ph}^{\phi=\pi}_S$ & $+$ & $+$ & $+$ & $+$ & $-$ & $-$ & $-$ & $-$
\end{tabular}

\vspace{10pt}

Note that the so-called \textit{phase} automorphism in the last table is a special case of the rotation
\begin{equation}
\textbf{Ph}^\phi : \Omega\longmapsto e^{i\phi}\Omega\,,\qquad \Upsilon\longmapsto e^{i\phi}\Upsilon\,,
\end{equation}
which was briefly pointed out in section~\ref{sec:TypeI}.

\vspace{5pt}

\subsubsection*{Odake $n=3$:}

$\qquad$\renewcommand\arraystretch{1.2}
\begin{tabular}{ r | c | c | c | c | c | c | c | c || c | c }
 & $T$ & $G$ & $G^3$ & $J$ & $A$ & $B$ & $C$ & $D$ & $N^1$ & $N^2$ \\\hline
Parity $\textbf{P}$ & $+$ & $-$ & $-$ & $+$ & $-$ & $-$ & $+$ & $+$ & $-$ & $-$ \\\hline
Mirror symmetry $\textbf{M}$ & $+$ & $+$ & $-$ & $-$ & $+$ & $-$ & $+$ & $-$ & $+$ & $-$ \\\hline
Phase $\textbf{Ph}^{\phi=\pi}$ & $+$ & $+$ & $+$ & $+$ & $-$ & $-$ & $-$ & $-$ & $-$ & $-$ 
\end{tabular}

\vspace{10pt}

\subsubsection*{$G_2$ SCA:}

$\qquad$\renewcommand\arraystretch{1.2}
\begin{tabular}{ r | c | c | c | c | c | c || c }
 & $T$ & $G$ & $\Phi$ & $X$ & $K$ & $M$ & $N$ \\\hline
Parity $\bm{\mathcal{P}}$ & $+$ & $-$ & $-$ & $+$ & $+$ & $-$ & $-$ \\\hline
GK mirror symmetry $\bm{\mathcal{M}}$ & $+$ & $+$ & $-$ & $+$ & $-$ & $+$ & $+$
\end{tabular}

\vspace{10pt}

The automorphism $\bm{\mathcal{M}}$ of the $G_2$ SCA was first observed in \cite{Becker:1996ay} (see also \cite{Roiban:2002iv}). It was interpreted as mirror symmetry of Joyce orbifolds by Gaberdiel and Kaste (GK) in \cite{Gaberdiel:2004vx}, as mentionned above.

\vspace{10pt}

We now point out a few relations between these automorphisms when the algebras are assembled together as per the TCS construction given in section~\ref{sec:Realisations}. Starting within region I, we notice that the T-duality map applied to the free algebra labelled by $\xi$, when combined with $\mathbf{Ph}^\pi$ on the $\text{Od}^{n=3}$ factor, induces an involution of $\text{Od}^{n=3}\times \text{Free}_\xi$. It can be checked explicitly with \eqref{eq:T7}--\eqref{eq:M7} that this involution restricts to a well-defined automorphism of the $G_2$ SCA sitting inside (c.f.\ \eqref{eq:InclusionI}). The map defined in this way is the  Gaberdiel--Kaste automorphism. We write this as
\begin{equation} \label{FOFautomT}
\mathbf{T}_\xi \circ \mathbf{Ph}^\pi \overset{\text{I}}{\dashrightarrow} \bm{\mathcal{M}} \,.
\end{equation}
Note that all the automorphisms in the tables above are involutive, so we can compose them in any order we want.

Another way to produce the Gaberdiel--Kaste automorphism in region $\text{I}$ is with a mirror automorphism of the Odake factor:
\begin{equation} \label{FOFautomM}
\mathbf{M} \circ \mathbf{Ph}^\pi \overset{\text{I}}{\dashrightarrow} \bm{\mathcal{M}} \,.
\end{equation}
Combining these two, we get the identity map
\begin{equation} \label{FOFautomM+T}
\mathbf{M} \circ \mathbf{T}_\xi \overset{\text{I}}{\dashrightarrow} \textbf{1} \,.
\end{equation}

For the realisation of $\text{Od}^{n=3}$ in region $\text{II}$ defined by the embedding $\text{II}_X$ in \eqref{eq:InclusionIICY}, we identify the following relations:
\begin{align}
\mathbf{M}_S \circ \mathbf{T}_\theta \circ \mathbf{Ph}^\pi_S &\overset{\text{II}_X}{\dashrightarrow} \mathbf{M} \,, \label{eq:automoM->M} \\
\mathbf{Ph}^\phi_S &\overset{\text{II}_X}{\dashrightarrow} \mathbf{Ph}^\phi \,. \label{eq:automoPh->Ph}
\end{align}

\subsection{Two TCS automorphisms} \label{sec:AutomTCS}

We finally discuss algebra automorphisms consistent all over the twisted connected sum geometry.

Starting from one side of the construction, say the $+$ side, we can perform for example the map \eqref{FOFautomM+T} locally in region $\text{I}^+$. At the level of the $G_2$ SCA, this acts as the identity despite being generated by non-trivial maps. The geometric interpretation is clearly a T-duality on the external $\mathbb{S}^1_\xi$ and mirror symmetry on the ACyl Calabi--Yau $X_+$.

The map $\mathbf{M}\circ\mathbf{T}_\xi$ from \eqref{FOFautomM+T} in region I$^+$, via the transition relations \eqref{eq:T3}--\eqref{eq:D3} described in section­~\ref{sec:Compatibility}, gives rise uniquely to an automorphism of the realisation in region $\text{II}^+$. As an example of how this is done, consider $J$ in region $\text{I}^+$, which becomes $\no{\psi_t \psi_\theta}+J_S$ in region $\text{II}^+$. For $J$ to change sign, we need $J_S$ to change sign, as well as either $\psi_t$ or $\psi_\theta$. Continuing in this way, we find how all generators of $\text{Od}^{n=2}\times (\text{Free})^3$ should transform. The result is
\begin{equation}
\mathbf{M}_S \circ \mathbf{Ph}^\pi_S \circ \mathbf{T}_\theta \circ \mathbf{T}_\xi \,.
\end{equation}
Geometrically, this corresponds to taking the mirror of the K3 surface $S_+$ and T-duals of the circles $\mathbb{S}^1_\theta$ and $\mathbb{S}^1_\xi$. Interestingly, a phase automorphism is also necessary.

Proceeding similarly as we progress from region $\text{II}^+$ to region $\text{II}^-$, now with the pasting automorphism $F$ in \eqref{eq:PastingAuto}, we find again a unique consistent map in $\text{II}^-$,
\begin{equation}
\mathbf{Ph}^\pi_S \circ \mathbf{M}_S \circ \mathbf{T}_\xi \circ \mathbf{T}_\theta \,.
\end{equation}
It is worth pointing out that the phase and mirror symmetry automorphisms in region $\text{II}^+$ get swapped as we get to region $\text{II}^-$. Finally, we reuse the asymptotic formulae \eqref{eq:T3}--\eqref{eq:D3} to reach a unique local form of the automorphism in region $\text{I}^-$.

Over the whole TCS construction, we have thus built a rigid arrangement of local maps starting from the choice \eqref{FOFautomM+T} in region $\text{I}^+$. It turns out that the automorphism generated on the $G_2$ SCA is the same in all regions. By the relations \eqref{FOFautomM+T} and \eqref{eq:automoM->M} it is easy to see that it is the identity. We summarise these results in table~\ref{tab:TCS_Id}.

\begin{table}[H]
\begin{center}
\renewcommand\arraystretch{1.5}
\begin{tabular}{ c | c | c | c | c }

Region & $\text{I}^+$ & $\text{II}^+$ & $\text{II}^-$ & $\text{I}^-$ \\\hline

Autom. of factors & \pbox{20cm}{$\mathbf{M}$ \\ $\mathbf{T}_\xi$} & \pbox{20cm}{~\vspace{8pt} \\ $\mathbf{M}_S$ \\ $\mathbf{Ph}^\pi_S$ \\ $\mathbf{T}_\theta$ \\ $\mathbf{T}_\xi$ \\} & \pbox{20cm}{~\vspace{8pt} \\ $\mathbf{Ph}^\pi_S$ \\ $\mathbf{M}_S$ \\ $\mathbf{T}_\xi$ \\ $\mathbf{T}_\theta$ \\} & \pbox{20cm}{~\vspace{-6pt} \\ $\mathbf{M}$ \\ $\mathbf{T}_\xi$} \\\hline

$G_2$ autom. & $\textbf{1}$ & $\textbf{1}$ & $\textbf{1}$ & $\textbf{1}$
\end{tabular}
\caption{TCS mirror automorphism $\mathcal{T}_4$}
\label{tab:TCS_Id}
\end{center}
\end{table}

By the geometrical interpretation given above, it seems clear that this scenario is a stringy version of the mirror map $\mathcal{T}_4$ defined in \cite{Braun:2017csz}. The authors of this paper choose for this map a special Lagrangian fibration of the ACyl Calabi--Yau manifolds and T-dualise along the $\mathbb{T}^3$ fibres and the external $\mathbb{S}^1_\xi$. As a consequence, the Calabi--Yau $2$- and $3$-folds get replaced by their mirrors by the SYZ argument \cite{Strominger:1996it}. We presently learn that this map $\mathcal{T}_4$ corresponds to the identity automorphism of the Shatashvili--Vafa SCA.\footnote{This fact has been verified explicitely for certain TCS obtained as Joyce orbifolds in \cite{Braun:2017csz}.}

Note that our point of view however does not rely on any SYZ fibration. Only the Calabi--Yau mirror automorphism enters our consideration. In fact, we need no geometric assumptions at all suggesting that $\mathcal{T}_4$ persists after stringy corrections. Another advantage of our approach is that we can systematically classify all possible combinations of automorphisms that are mutually consistent in the TCS realisation and lead to global $G_2$ automorphisms. We have not pursed this in full generality, but we can immediately identify a second map besides $\mathcal{T}_4$.

Instead of \eqref{FOFautomM+T}, let us now start with \eqref{FOFautomM} in region $\text{I}^+$. Consistency with the transition maps again allows to progress inside the geometry and through the pasting isomorphism to produce the results of table~\ref{tab:TCS_M}. (This is the only possibility consistent with \eqref{FOFautomM} in region $\text{I}^+$.)

\begin{table}[H]
\begin{center}
\renewcommand\arraystretch{1.5}
\begin{tabular}{ c | c | c | c | c }

Region & I$^+$ & II$^+$ & II$^-$ & I$^-$ \\\hline

Autom. of factors & \pbox{20cm}{$\mathbf{M}$ \\ $\mathbf{Ph}^\pi$} & \pbox{20cm}{~\vspace{8pt} \\ $\mathbf{M}_S$ \\ $\mathbf{T}_\theta$ \\} & \pbox{20cm}{~\vspace{8pt} \\ $\mathbf{Ph}^\pi_S$ \\ $\mathbf{T}_\xi$ \\} & \pbox{20cm}{~\vspace{-2pt} \\ $\mathbf{Ph}^\pi$ \\ $\mathbf{T}_\xi$} \\\hline

$G_2$ autom. & $\bm{\mathcal{M}}$ & $\bm{\mathcal{M}}$ & $\bm{\mathcal{M}}$ & $\bm{\mathcal{M}}$
\end{tabular}
\caption{TCS mirror automorphism $\mathcal{T}_3$}
\label{tab:TCS_M}
\end{center}
\end{table}

The geometric interpretation is now much different. Mirror symmetry only acts on one side of the construction (here $X_+$ and its asymptotic cross-section $S_+\times \mathbb{S}^1_\theta$), while T-duality acts on the other side (along with a phase shift). The induced $G_2$ automorphism is again identical over the whole geometry, but this time it is the Gaberdiel--Kaste mirror map $\bm{\mathcal{M}}$.

This case corresponds to the mirror map $\mathcal{T}_3$ from \cite{Braun:2017csz}. Geometrically, only the $+$ side ACyl Calabi--Yau was assumed to be SYZ fibered. On the other side, the K3 surface $S_-$ was assumed to be elliplically fibered.

\section{Discussion} \label{sec:Conclusion}

Several parallels exist between the geometry of Riemannian manifolds admitting parallel spinors and the conformal field theories describing string compactifications on them. Yet this correspondence is tenuous. For example, it was recently established \cite{Melnikov:2017yvz,Fiset:2017auc,delaOssa:2018azc} that the Shatashvili--Vafa SCA accounts for heterotic string compactifications on $G_2$-structure manifolds with torsion and vector bundles over them; a radical departure from the original set up of $G_2$ holonomy and type II strings. A fundamentual source of ambiguity is the generalised mirror conjecture of \cite{Shatashvili:1994zw}: the deficiency of the conformal field theory to decipher aspects of the target manifold is precisely explained by the existence of multiple (mirror) geometries corresponding to the same conformal field theory (at least up to marginal deformations and automorphisms). 

In spite of this, we have explained in this paper how to take advantage of relations like \eqref{eq:GeometryVsSCA} and inspiration from already existing realisations of the Shatashvili--Vafa $G_2$ SCA \cite{Shatashvili:1994zw,Figueroa-OFarrill:1996tnk} to establish a realisation for twisted connected sums. Our realisation could perhaps seem unsurprising based on the mathematically established existence of Ricci-flat metrics on TCSs, which suggests a fixed point of the RG flow. However, we should recall that a true CFT corresponds to a Ricci-flat target manifold only to leading orders in worldsheet perturbation theory. Moreover, the aforementioned ambiguities in the correspondence between CFT and geometry makes worthwhile an explicit construction.

Our approach gives a certain exhaustivity in the identification of mirror maps for TCS geometries. A preliminary analysis was sufficient to recover a stringy version of both mirror maps $\mathcal{T}_4$ and $\mathcal{T}_3$ suggested by Braun and Del Zotto \cite{Braun:2017csz}. It is quite possible that a more systematic inspection of the automorphisms of the algebras we used --- $G_2$ SCA, Odake $n=3$ and Odake $n=2$ --- would reveal new mirror maps by following the logic of section~\ref{sec:AutomTCS}.

Perhaps the most interesting future direction to pursue with our TCS realisation is the highest weight representation theory of these symmetry algebras. Interesting lessons about mirror symmetry could be learned from a firmer grasp on the Hilbert space of the CFT. A prime example is to determine the effect of the mirror maps $\mathcal{T}_3$ and $\mathcal{T}_4$ on the Betti numbers of the TCS manifold. This would be in analogy with the exchange of $h^{1,1}$ and $h^{2,1}$ in Calabi--Yau mirror symmetry, which follows from the automorphism \eqref{eq:Calabi-YauMiSYAut}.

The key step for this task is the correspondence between Ramond--Ramond ground states and cohomology groups explained in \cite{Shatashvili:1994zw}. This correspondence remains ambiguous, partly due to the generalised mirror conjecture stated above. In the case of the $G_2$ SCA, only the sum
\begin{equation}
b^2+b^3
\end{equation}
of Betti numbers can be obtained from the algebra; not the indvidual values. We can suggest however that the explicit realisation of the $G_2$ SCA we have given will allow a more detailed understanding of the ground states and, in turn, of the Betti numbers.

Carrying out this program explicitly requires the representation theory of the local conformal algebras used in regions $\text{I}$ and $\text{II}$. In spite of the known representation theory of the Odake and free algebras, this is technically complicated because adding generators substantially alters the labelling of highest weight states and the creation operators, and thus the whole tower of descendants. Subsequently, one would have to devise transition maps for the ground states following the philosophy of section~\ref{sec:Compatibility} and understand how to make global statements about the topology. Comparison with the geometric approach based on a Mayer--Vietoris sequence \cite{Corti:2012kd} would certainly be useful for this.

We should admit the possibility that our presentation in this paper of the twisted connected sum realisation of the $G_2$ CFT is perhaps not the best way to interpret our result. We trust however that it promises a better control on twisted connected sum manifolds from a worldsheet angle. A clearer view would certainly arise from a specific $G_2$ CFT built on the principles of this paper. We may for example try to use Gepner models for all Calabi--Yau factors in the TCS construction and assemble them together following the ideas in section~\ref{sec:Compatibility}.\footnote{This suggestion was made by S. Sch\"afer-Nameki.} This could yield a Gepner-type model for $G_2$ manifolds in the spirit of the discussion in \cite{Roiban:2001cp,Roiban:2002iv} (see also \cite{Eguchi:2001ip, Eguchi:2003yy,  Blumenhagen:2001jb, Noyvert:2002mc, Sugiyama:2001qh, Sugiyama:2002ag}).

Finally, we wish to point out the recent work \cite{Braun:2018joh} where a construction of $Spin(7)$ holonomy manifolds is given based on the twisted connected sum principle. We expect there to exist a corresponding realisation of the $Spin(7)$ superconformal algebra, also due to Shatashvili and Vafa \cite{Shatashvili:1994zw}. Mirror maps for these geometries would be an interesting by-product.

\section*{Acknowledgement}

Andreas Braun deserves special thanks for suggesting to investigate twisted connected sum mirror maps from a conformal field theoretic point of view. I am indebted to him, Xenia de la Ossa, Christopher Beem, and Sakura Sch\"afer-Nameki for their useful comments on previous drafts, their guidance and their support along the way to completion of this project. I also benefited from discussions with Sebastjan Cizel and Max H\"ubner and I acknowledge Kris Thielemans' \textit{Mathematica}\textsuperscript{®} package for computing operator product expansions. 
My research is financed by a Reidler scholarship from the Mathematical Institute at the University of Oxford and by a FRQNT doctoral scholarship from the Government of Quebec.

\appendix

\section{Superconformal algebras} \label{app:algebras}

This appendix clarifies some topics in superconformal algebras that are relevant to the main text. We also detail our normalisations and the explicit operator product expansions used in our computations.

We refer to the algebraic structures discussed in this note as \textit{superconformal algebras}, or SCAs for short, but they are known under multiple other names in the literature. Some are more familiar to the mathematical audience --- such as ``vertex (operator) algebras'' --- while in physics they are also called ``chiral'', ``current'', or ``$\mathcal{W}$''-algebras. They are central to the abstract description of $2$-dimensional conformal field theories based on their symmetries, in which case a pair of SCAs are needed; one for each chirality.

Essentially here, we have in mind a finite set of \textit{generators} $\{A(z),B(z),\ldots\}$ (interchangeably called ``currents'', ``operators'', or ``fields'') which are meromorphic on a subset of $\mathbb{C}$. The SCA is specified by \textit{operator product expansions} (OPE) relating pairwise these generators. For general reviews, see \cite{Bouwknegt:1992wg,Kac1998VOA,FrenkelBZ2001} and \cite{Thielemans:1994er}. In particular the last reference details all the conventions used by Thielemans in his \textit{OPEdefs} \textit{Mathematica}\textsuperscript{®} package \cite{Thielemans:1991uw}, which was used for calculating OPEs for the present work.


There is a slight technicality that some SCAs suffer from and we address it here because it affects the algebras we are interested in. The most constraining axiom behind a consistent SCA is that of associativity. In physics terms, it guarantees that the order in which contractions are made within correlators does not affect the final result (crossing symmetry). At the level of OPEs, it translates into Jacobi-like identities between triplets of fields. A useful consequence of associativity is the formula
\begin{equation}
\wick{ \c A(z) \, \no{\c {BC}}(w)} = \frac{1}{2\pi i} \oint_{\mathbb{S}^1(w)} \frac{\dd x}{x-w}\left[ \wick{ \c A(z) \c B(x)} C(w) + (-1)^{|A||B|} B(x) \wick{ \c A(z) \c C(w)} \right]\,,
\end{equation}
where $|A|=\pm 1$ is the parity of $A$.

Sometimes, such relations are satisfied on the nose by the SCA. Generically however, one must allow so-called \textit{null} fields $\{N^1,N^2,\ldots\}$ as well as the ordinary generators. Physically, they are characterised by their vanishing correlators with all other fields. In particular, null fields could arise in any consistency condition for the SCA without impacting measurable quantities. The infinite set of fields obtained from these null fields by taking any combinations of derivatives, normal ordered products and OPEs with any other fields is called the \textit{ideal} generated by $\{N^1,N^2,\ldots\}$. We denote it by $\left\langle N^1, N^2,\ldots\right\rangle$. The references mentioned above provide more details about null ideals. We give in the main text some more illustration of their importance.

\subsection{Free boson and fermion} \label{app:free}

Let $\xi$ be a real boson and $\psi$, a real chiral fermion. Let $j=i\xi'$. The free OPE relations are
\begin{align}
\wick{ \c j(z) \c j(w)} &=\frac{1}{(z-w)^2}\,,\\
\wick{ \c \psi(z) \c \psi(w)} &=\frac{1}{z-w}\,.
\end{align}

\subsection{$\mathcal{N}=1$ Virasoro} \label{app:N=1Vir}

The generators of $\mathcal{N}=1$ Virasoro are $T$, the energy-momentum tensor, and $G$, the supersymmetry current. They are both real and form a multiplet $(G,2T)$. They obey the following algebra:
\begin{equation}
\wick{\c T(z) \c T(w)}=\frac{c/2}{(z-w)^4}+\frac{2T(w)}{(z-w)^2}+\frac{T'(w)}{z-w}\,,
\end{equation}
\begin{equation}
\wick{\c T(z) \c G(w)} =\frac{3/2}{(z-w)^2}G(w)+\frac{G'(w)}{z-w}\,,
\end{equation}
\begin{equation}
\wick{\c G(z) \c G(w)} =\frac{2c/3}{(z-w)^3}+\frac{2T(w)}{z-w}\,.
\end{equation}

\subsection{$\mathcal{N}=2$ Virasoro} \label{app:N=2Vir}

In addition to the $\mathcal{N}=1$ generators $(G=G^{0},2T)$, $\mathcal{N}=2$ Virasoro has the real multiplet $(J,G^{3})$. It is composed of a second supersymmetry current $G^{3}$ and its superpartner $J$, which is a $U(1)$ current. In addition to the $\mathcal{N}=1$ OPEs we have the following relations.

\begin{equation}
\wick{\c T(z) \c J(w)} =\frac{J(w)}{(z-w)^2}+\frac{J'(w)}{z-w}
\end{equation}
\begin{equation}
\wick{\c T(z)  \c G^{3}(w)}=\frac{3/2}{(z-w)^2}G^{3}(w)+\frac{(G^3)'(w)}{z-w}
\end{equation}
\begin{equation}
\wick{\c G^{0}(z) \c J(w)}=\frac{G^{3}(w)}{z-w}
\end{equation}
\begin{equation}
\wick{\c G^{0}(z) \c G^{3}(w)}=\frac{2J(w)}{(z-w)^2}+\frac{J'(w)}{z-w}
\end{equation}
\begin{equation}
\wick{\c G^{3}(z) \c G^{3}(w)}=\frac{2c/3}{(z-w)^3}+\frac{2T(w)}{z-w}
\end{equation}
\begin{equation}
\wick{\c J(z) \c J(w)}=-\frac{c/3}{(z-w)^2}
\end{equation}
\begin{equation}
\wick{\c G^{3}(z) \c J(w)}=-\frac{G^{0}(w)}{z-w}
\end{equation}

Another choice of basis for the generators is common in the literature; see e.g.  \cite{Odake:1988bh,Quigley:2014rya,Alim:2012gq,Greene:1996cy}. We render the $U(1)$ current imaginary by defining
\begin{equation}
I=-iJ
\end{equation}
and we combine the supercurrents as
\begin{equation}
G^{\pm}=\frac{1}{\sqrt{2}}\left(G^{0}\pm iG^{3}\right).
\end{equation}
Note that $G^+$ and $G^-$ are complex conjugate to each other: $G^-=(G^+)^*$. With these definitions, the OPEs that do not involve $T$ are as follows.

\begin{equation}
\wick{\c I(z) \c I(w)}=\frac{c/3}{(z-w)^2}
\end{equation}
\begin{equation}
\wick{\c I(z) \c G^{\pm}(w)}=\pm\frac{G^{\pm}(w)}{z-w}
\end{equation}
\begin{equation}
\wick{\c G^+(z) \c G^-(w)}=\frac{2c/3}{(z-w)^3}+\frac{2I(w)}{(z-w)^2}+\frac{(I'+2T)(w)}{z-w}
\end{equation}
\begin{equation}
\wick{\c G^\pm(z) \c G^\pm(w)}=0
\end{equation}

\subsection{Odake} \label{app:Odake}

The Odake algebra \cite{Odake:1988bh} with central charge $c=3n$, $n\in\mathbb{N}$, introduces four new real generators or, equivalently, two complex generators:
\begin{equation}
\Omega=A+iB \qquad \text{and}\qquad \Upsilon=\frac{C+iD}{\sqrt{2}}\,.
\end{equation}
Their complex conjugates are denoted with the star; e.g.\ $\Omega^*=A-iB$. We start by providing the OPEs with the $\mathcal{N}=2$ generators. Later we will consider individually $n=2$ and $n=3$ and give the OPEs involving only $\Omega$, $\Upsilon$ and their conjugates.

The field $\Omega$ is a weight-$n/2$ primary with respect to $T$ and it has $U(1)$ charge $n$:
\begin{equation}
\wick{\c T(z) \c \Omega(w)}=\frac{n/2}{(z-w)^2}\Omega(w)+\frac{\Omega'(w)}{z-w}\,,
\end{equation}
\begin{equation}
\wick{\c I(z) \c \Omega(w)}=\frac{n\Omega(w)}{z-w}\,.
\end{equation}
The field $\Upsilon$ has weight $\tfrac{n+1}{2}$ and is primary with respect to $T$. It has $U(1)$ charge $n-1$:
\begin{equation}
\wick{\c T(z) \c \Upsilon(w)}=\frac{(n+1)/2}{(z-w)^2}\Upsilon(w)+\frac{\Upsilon'(w)}{z-w}\,,
\end{equation}
\begin{equation}
\wick{\c I(z) \c \Upsilon(w)}=\frac{n-1}{z-w}\Upsilon(w)\,.
\end{equation}
$\Omega$ and $\Upsilon$ are related by supersymmetry as follows.
\begin{equation}
\wick{\c G^+(z) \c \Omega(w)}=0
\end{equation}
\begin{equation}
\wick{\c G^-(z) \c \Omega(w)}=\frac{2\Upsilon(w)}{z-w}
\end{equation}
\begin{equation}
\wick{\c  G^+(z) \c \Upsilon(w)}=\frac{n\Omega(w)}{(z-w)^2}+\frac{\Omega'(w)}{z-w}
\end{equation}
\begin{equation}
\wick{\c  G^-(z) \c \Upsilon(w)}=0
\end{equation}

\subsubsection{Odake $n=2$} \label{app:Odake2}

We now discuss the OPEs involving $\Omega$, $\Upsilon$, and their conjugates in the case $n=2$. There are a priori $10$ remaining OPEs. In real basis, they are $AA$, $AB$, $AC$, $AD$, $BB$, $BC$, $BD$, $CC$, $CD$, and $DD$. In fact, associativity of the OPE can be used to deduce them all in terms of the first three, for instance. We thus start with these and take
\begin{equation}\label{eq:AA}
\wick{\c A(z) \c A(w)}=-\frac{2}{(z-w)^2}\,,
\end{equation}
\begin{equation}
\wick{\c B(z) \c B(w)}=-\frac{2}{(z-w)^2}\,,
\end{equation}
\begin{equation}\label{eq:AB}
\wick{\c A(z) \c B(w)}=-\frac{2J(w)}{z-w}\,.
\end{equation}

As it turns out, the Odake $n=2$ algebra is the small $\mathcal{N}=4$ Virasoro in disguise \cite{Odake:1988bh,Ali:1999ut,Ademollo:1976wv}. This algebra has four superymmetry currents $G^{0}$, $G^{1}$, $G^{2}$, $G^{3}$ and, as compared to the basic multiplet $(G^{0},2T)$, each new supercurrent introduces a new $U(1)$ current: $\mathcal{J}^{1}$, $\mathcal{J}^{2}$, $\mathcal{J}^{3}$, where $\mathcal{J}^{3}$ is proportional to the generator that we called $J$ previously. These currents form an $SU(2)$ Kac-Moody algebra at level $k=c/6=1$ \cite{Ali:1999ut}:
\begin{equation}
\wick{\c {\mathcal{J}}^i(z) \c {\mathcal{J}}^j(w)}=\frac{k/2}{(z-w)^2}\delta_{ij}+\frac{i\epsilon_{ijk}\mathcal{J}^k(w)}{z-w}\,.
\end{equation}

The OPEs given in \eqref{eq:AA}--\eqref{eq:AB} are in fact consistent with the following identifications:
\begin{equation}
\mathcal{J}^{1}=\frac{A}{2i}\,,\qquad \mathcal{J}^{2}=\frac{B}{2i}\,,\qquad \mathcal{J}^{3}=\frac{J}{2i}\,.
\end{equation}
As for $\Upsilon$, it is clear by dimensional analysis, that it is related to the two new supersymmetry currents $G^1$, $G^2$.
From this, associativity is enough to fix the remaining OPEs:
\begin{equation}
\wick{\c A(z) \c C(w)}=\frac{G^0(w)}{z-w}\,,
\end{equation}
\begin{equation}
\wick{\c A(z) \c D(w)}=-\frac{G^3(w)}{z-w}\,,
\end{equation}
\begin{equation}
\wick{\c B(z) \c C(w)}=\frac{G^3(w)}{z-w}\,,
\end{equation}
\begin{equation}
\wick{\c B(z) \c D(w)}=\frac{G^0(w)}{z-w}\,,
\end{equation}
\begin{equation}
\wick{\c C(z) \c C(w)}=\frac{4}{(z-w)^3}+\frac{2T(w)}{z-w}\,,
\end{equation}
\begin{equation}
\wick{\c C(z) \c D(w)}=\frac{2J}{(z-w)^3}+\frac{J'(w)}{z-w}\,,
\end{equation}
\begin{equation}
\wick{\c D(z) \c D(w)}=\frac{4}{(z-w)^3}+\frac{2T(w)}{z-w}\,.
\end{equation}

\subsubsection{Odake $n=3$} \label{app:Odake3}

\begin{equation}
\wick{\c A(z) \c A(w)}=-\frac{4}{(z-w)^3}+\frac{2\no{JJ}(w)}{z-w}
\end{equation}
\begin{equation}
\wick{\c B(z) \c B(w)}=-\frac{4}{(z-w)^3}+\frac{2\no{JJ}(w)}{z-w}
\end{equation}
\begin{equation}
\wick{\c A(z) \c B(w)}=-\frac{4J(w)}{(z-w)^2}-\frac{2J'(w)}{z-w}
\end{equation}
\begin{equation}
\wick{\c A(z) \c C(w)}=-\frac{2G^0(w)}{(z-w)^2}-\frac{2\no{JG^3}(w)}{z-w}
\end{equation}
\begin{equation}
\wick{\c A(z) \c D(w)}=\frac{2G^3(w)}{(z-w)^2}-\frac{2\no{JG^0}(w)}{z-w}
\end{equation}
\begin{equation}
\wick{\c B(z) \c C(w)}=-\frac{2G^3(w)}{(z-w)^2}+\frac{2\no{JG^0}(w)}{z-w}
\end{equation}
\begin{equation}
\wick{\c B(z) \c D(w)}=-\frac{2G^0(w)}{(z-w)^2}-\frac{2\no{JG^3}(w)}{z-w} 
\end{equation}
\begin{equation}
\wick{\c C(z) \c C(w)}=-\frac{12}{(z-w)^4}+\frac{(2\no{JJ}-4T)(w)}{(z-w)^2}+\frac{(2\no{J'J}-2T')(w)}{z-w}
\end{equation}
\begin{equation}
\wick{\c C(z) \c D(w)}=-\frac{8J(w)}{(z-w)^3}-\frac{4J'(w)}{(z-w)^2}+\frac{(2\no{GG^3}-4\no{TJ})(w)}{z-w}
\end{equation}
\begin{equation}
\wick{\c D(z) \c D(w)}=-\frac{12}{(z-w)^4}+\frac{(2\no{JJ}-4T)(w)}{(z-w)^2}+\frac{(2\no{J'J}-2T')(w)}{z-w}
\end{equation}

Associativity of the OPE is only realised modulo the ideal generated by the null weight-$5/2$ fields \cite{Odake:1988bh}
\begin{equation}
N^1=A' -\no{JB}\,,\qquad N^2=B' +\no{JA}\,.
\end{equation}

\subsection{$G_2$ algebra} \label{app:SV}

The Shatashvili--Vafa $G_2$ superconformal algebra is an extension of $\mathcal{N}=1$ Virasoro $(G,2T)$ with central charge $c=21/2$. The extra multiplets are $(\Phi,K)$ and $(X,M)$ where the lowest components have respectively weight $3/2$ and $2$. We start with the OPEs with the $\mathcal{N}=1$ generators.

\begin{equation}
\wick{\c T(z) \c \Phi(w)}=\frac{3/2}{(z-w)^2}\Phi(w)+\frac{\Phi'(w)}{z-w}
\end{equation}
\begin{equation}
\wick{\c T(z) \c X(w)}=-\frac{7/4}{(z-w)^4}+\frac{2X(w)}{(z-w)^2}+\frac{X'(w)}{z-w}
\end{equation}
\begin{equation}
\wick{\c T(z) \c K(w)}=\frac{2K(w)}{(z-w)^2}+\frac{K'(w)}{z-w}
\end{equation}
\begin{equation}
\wick{\c T(z) \c M(w)}=-\frac{1/2}{(z-w)^3}G(w)+\frac{5/2}{(z-w)^2}M(w)+\frac{M'(w)}{z-w}
\end{equation}
\begin{equation}
\wick{\c G(z) \c \Phi(w)}=\frac{K(w)}{z-w}
\end{equation}
\begin{equation}
\wick{\c G(z) \c X(w)}=-\frac{1/2}{(z-w)^2}G(w)+\frac{M(w)}{z-w}
\end{equation}
\begin{equation}
\wick{\c G(z) \c K(w)}=\frac{3\Phi(w)}{(z-w)^2}+\frac{\Phi'(w)}{z-w}
\end{equation}
\begin{equation}
\wick{\c G(z) \c M(w)}=-\frac{7/2}{(z-w)^4}+\frac{\left(T+4X\right)(w)}{(z-w)^2}+\frac{X'(w)}{z-w}
\end{equation}

We now provide the OPEs of the new generators with each other.

\begin{equation}
\wick{\c \Phi(z) \c \Phi(w)}=-\frac{7}{(z-w)^3}+\frac{6X(w)}{z-w}
\end{equation}
\begin{equation}
\wick{\c \Phi(z) \c X(w)}=-\frac{15/2}{(z-w)^2}\Phi(w)-\frac{5/2}{(z-w)}\Phi'(w)
\end{equation}
\begin{equation}
\wick{\c \Phi(z) \c K(w)}=-\frac{3G(w)}{(z-w)^2}-\frac{3}{z-w}\left(M+\frac{1}{2}G'\right)(w)
\end{equation}
\begin{equation}
\wick{\c \Phi(z) \c M(w)}=\frac{9/2}{(z-w)^2}K(w)+\frac{1}{z-w}\left(3\no{\Phi G}-\frac{1}{2} K'\right)(w)
\end{equation}
\begin{equation}
\wick{\c X(z) \c X(w)}=\frac{35/4}{(z-w)^4}-\frac{10X(w)}{(z-w)^2}-\frac{5X'(w)}{z-w}
\end{equation}
\begin{equation}
\wick{\c X(z) \c K(w)}=-\frac{3K(w)}{(z-w)^2}-\frac{3\no{\Phi G}(w)}{z-w}
\end{equation}
\begin{equation}
\wick{\c X(z) \c M(w)}=-\frac{9/2}{(z-w)^3}G(w)-\frac{1}{(z-w)^2}\left(5M+\frac{9}{4}G'\right)(w)+\frac{1}{z-w}\left(4\no{XG}+\frac{1}{2}M'+\frac{1}{4} G''\right)(w)
\end{equation}
\begin{equation}
\wick{\c K(z) \c K(w)}=-\frac{21}{(z-w)^4}+\frac{6(X-T)(w)}{(z-w)^2}+\frac{3(X'-T')(w)}{z-w}
\end{equation}
\begin{equation}
\wick{\c K(z) \c M(w)}=-\frac{15\Phi(w)}{(z-w)^3}-\frac{11/2}{(z-w)^2}\Phi'(w)+\frac{\left(3\no{GK}-6\no{TP}\right)(w)}{z-w}
\end{equation}
\begin{align}
\wick{\c M(z) \c M(w)}=-\frac{35}{(z-w)^5}&+\frac{(20X-9T)(w)}{(z-w)^3}+\frac{1}{(z-w)^2}\left(10X'-\frac{9}{2}T'\right)(w) \\
&+\frac{1}{z-w}\left(\frac{3}{2}(X''-T'')-4\no{GM}+8\no{TX}\right)(w) \nonumber
\end{align}

Associativity of the OPE is only realised modulo the ideal generated by the null weight-$5/2$ field \cite{Figueroa-OFarrill:1996tnk}
\begin{equation}
N=4\no{GX}-2\no{\Phi K}-4M'-G''\,.
\end{equation}


\normalem


\bibliographystyle{JHEP}

\bibliography{FisetSEP2018}

\end{document}